\documentclass[aps,prd,preprint,preprintnumbers,showpacs,nofootinbib]{revtex4}
\usepackage{graphicx}
\usepackage{amsmath}
\usepackage{amssymb}
\usepackage{amsfonts}
\usepackage{bm}
\usepackage{color}

\def\be{\begin{equation}}
\def\ee{\end{equation}}
\def\bea{\begin{eqnarray}}
\def\eea{\end{eqnarray}}

\begin{document}

\title{Finite temperature effects in Bose-Einstein Condensed dark matter
halos}
\author{Tiberiu Harko}
\email{harko@hkucc.hku.hk}
\affiliation{Department of Physics and Center for Theoretical and Computational Physics,
The University of Hong Kong, Pok Fu Lam Road, Hong Kong, P. R. China}
\author{Enik\H{o} J. M. Madarassy}
\email{eniko.madarassy@physics.uu.se}
\affiliation{Division of Astronomy and Space Physics, Uppsala University, 751 20 Uppsala, Sweden}

\begin{abstract}
Once the critical temperature of a cosmological boson gas is less than the critical temperature, a Bose-Einstein Condensation process can always take place during the cosmic history of the universe. Zero temperature condensed dark matter can be described as a non-relativistic, Newtonian gravitational condensate, whose density and pressure are related by a barotropic equation of state, with barotropic index equal to one.  In the present paper we analyze the effects of the finite dark matter temperature on the properties of the dark matter halos. We formulate the basic equations describing the finite temperature condensate, representing a generalized Gross-Pitaevskii equation that takes into account the presence of the thermal cloud. The static condensate and thermal cloud in thermodynamic equilibrium is analyzed in detail, by using the Hartree-Fock-Bogoliubov and Thomas-Fermi approximations. The condensed dark matter and thermal cloud density and mass profiles at finite temperatures are explicitly obtained. Our results show that when the temperature of the condensate and of the thermal cloud are much smaller than the critical Bose-Einstein transition temperature, the zero temperature density and mass profiles give an excellent description of the dark matter halos. However, finite temperature effects may play an important role in the early stages of the cosmological evolution of the dark matter condensates.

\end{abstract}

\pacs{67.85.Jk, 04.40.Dg, 95.30.Cq, 95.30.Sf}
\maketitle

\section{Introduction}

At very low temperatures, all particles in a dilute Bose gas condense to the
same quantum ground state, forming a Bose-Einstein Condensate (BEC), i.e., a
sharp peak over a broader distribution in both coordinates and momentum
space. Particles become correlated with each other when their wavelengths
overlap, that is, the thermal wavelength $\lambda _{T}$ is greater than the
mean inter-particles distance $l$. This happens at a temperature $T<2\pi
\hbar ^{2}/mk_{B}n^{2/3}$, where $m$ is the mass of the particle in the
condensate, $n$ is the number density, and $k_{B}$ is Boltzmann's constant
\cite{Da99,Pe}. A coherent state develops when the particle density is enough
high, or the temperature is sufficiently low. Quantum degenerate gases have
been created by a combination of laser and evaporative cooling techniques,
opening several new lines of research, at the border of atomic, statistical
and condensed matter physics \cite{Da99,rev,Pe}.

The Bose-Einstein condensation process was first observed experimentally in
1995 in dilute alkali gases, such as vapors of rubidium and sodium, confined
in a magnetic trap, and cooled to very low temperatures. A sharp peak in the
velocity distribution was observed below a critical temperature, indicating
that condensation has occurred, with the alkali atoms condensed in the same
ground state and showing a narrow peak in the momentum space and in the
coordinate space \cite{exp}. The Bose-Einstein condensation of hydrogen atoms \cite{Hy}, of calcium atoms \cite{Ca}, and of strontium atoms \cite{Sr} have also been observed.

Cosmological data provide compelling evidence that about 95\% of the
content of the Universe resides in two unknown forms of energy that we call
dark matter and dark energy: the first residing in bound objects as
non-luminous matter \cite{dm}, the latter in the form of a zero-point energy
that pervades the whole Universe \cite{PeRa03}. The dark matter is thought
to be composed of cold neutral weakly interacting massive particles, beyond
those existing in the Standard Model of Particle Physics, and not yet
detected in accelerators or in dedicated direct and indirect searches. There
are many possible candidates for dark matter, the most popular ones being
the axions and the weakly interacting massive particles (WIMP) (for a review
of the particle physics aspects of dark matter see \cite{OvWe04}). Their
interaction cross section with normal baryonic matter, while extremely
small, are expected to be non-zero and we may expect to detect them
directly. Scalar fields or other long range
coherent fields coupled to gravity have also intensively been used to model
galactic dark matter \cite{scal}. 

At galactic scales $\sim 10$ kpc, the $%
\Lambda $CDM model meets with severe difficulties in explaining the observed
distribution of the invisible matter around the luminous one. In fact, $N$%
-body simulations, performed in this scenario, predict that bound halos
surrounding galaxies must have very characteristic density profiles that
feature a well pronounced central cusp, $\rho _{NFW}(r)=\rho
_{s}/(r/r_{s})(1+r/r_{s})^{2}$ \cite{nfw}, where $r_{s}$ is a scale radius
and $\rho _{s}$ is a characteristic density. On the observational side,
high-resolution rotation curves show, instead, that the actual distribution
of dark matter is much shallower than the above, and it presents a nearly
constant density core: $\rho _{B}(r)=\rho
_{0}r_{0}^{3}/(r+r_{0})(r^{2}+r_{0}^{2})$ \cite{bur}, where $r_{0}$ is the
core radius and $\rho _{0}$ is the central density.

The possibility that the dark matter, which is required to explain the
dynamics of the neutral hydrogen clouds at large distances from the galactic
center, could be in the form of a Bose-Einstein condensate was proposed
initially in \cite{Sin}. To study the condensate the non-relativistic
Gross-Pitaevskii equation was investigated numerically. An alternative
approach was introduced in \cite{BoHa07}. By introducing the Madelung
representation of the wave function, the dynamics of the system can be
formulated in terms of the continuity equation and of the hydrodynamic Euler
equations. Hence dark matter can be described as a non-relativistic,
Newtonian Bose-Einstein gravitational condensate gas, whose density and
pressure are related by a barotropic equation of state. In the case of a
condensate with quartic non-linearity, the equation of state is polytropic
with index $n=1$. To test the validity of the model the Newtonian tangential
velocity equation of the model was fitted with a sample of rotation curves
of low surface brightness and dwarf galaxies, respectively. A very good
agreement was found between the theoretical rotation curves and the
observational data for the low surface brightness galaxies. Therefore dark
matter halos can be described as an assembly of light individual bosons that
acquire a repulsive interaction by occupying the same ground energy state.
That prevents gravity from forming the central density cusps. The condensate
particle is light enough to naturally form condensates of very small masses
that later may coalesce, forming the structures of the Universe in a similar
way than the hierarchical clustering of the bottom-up CDM picture. Then, at
large scales, BEC perfectly mimic an ensemble of cold particles, while at
small scales quantum mechanics drives the mass distribution. Different cosmological and astrophysical properties of condensed dark matter halos have been extensively investigated \cite{BEC,Har, Har1}.

All of the previous researches have been done by assuming that the condensed dark matter is at zero temperature. This assumptions is certainly a very good approximation for the description of dark matter in thermodynamic equilibrium with the cosmic microwave background, and for the analysis of the galactic rotation curves. It is already well established in condensed matter physics that the zero-temperature Gross-Pitaevskii equation gives an excellent quantitative descriptions of the Bose-Einstein condensates for $T\leq 0.5T_{BEC}$, where $T_{BEC}$ is the Bose-Einstein transition temperature. This condition is obviously satisfied by the dark matter halos of the low-redshift galaxies. However, in the early Universe, immediately after the condensation, finite temperature effects could have played an important role, and significantly affect the cosmological evolution.

It is the purpose of the present paper to study the finite temperature properties of the
gravitationally self-bound Bose-Einstein dark matter condensates of
collisionless particles. In particular, we focus on the description and the static properties of
the condensates interacting with a thermal cloud. The basic equations describing the gravitationally trapped condensate at finite temperature are obtained, and their static solution, describing a condensate in thermal equilibrium with a thermal cloud, is obtained by using the Hartree-Fock-Bogoliubov and Thomas-Fermi approximations, respectively. The temperature dependent dark matter density and mass profiles are obtained, as well the (temperature dependent) radius of the condensate and thermal cloud system. By using the explicit form of the physical parameters it immediately follows that the finite temperature effects do not play any significant role in the characterization  of the physical properties of the dark matter halos at low  redshifts, for which the static solutions of the standard Gross-Pitaevskii equation \cite{Da99,Pe} give an excellent description.

The present paper is organized as follows. The basic equations describing finite temperature Bose-Einstein condensates are presented in Section \ref{2}. The effects of the finite temperature on the dark matter halo profiles, trapped by their gravitational potential, are presented in Section \ref{3_s}. We discuss and conclude our results in Section \ref{4}.

\section{Bose-Einstein condensates at finite temperature}\label{2}

The Heisenberg equation of motion for the quantum field operator $\hat{\Phi}$
describing the dynamics of a Bose-Einstein condensate at arbitrary
temperatures is given by \cite{Za99, Prouk, Gr}
\begin{equation}
i\hbar \frac{\partial \hat{\Phi}\left( t,\vec{r}\right) }{\partial t}=\left[
-\frac{\hbar ^{2}}{2m}\Delta +mV_{grav}\left( \vec{r}\right) +g^{\prime }\hat{%
\Phi}^{+}\left( t,\vec{r}\right) \hat{\Phi}\left( t,\vec{r}\right) \right]
\hat{\Phi}\left( t,\vec{r}\right) ,  \label{1}
\end{equation}
where $m$ is the mass of the condensed particle, $V_{grav}\left( \vec{r}%
\right) $ is the gravitational trapping potential, and $g^{\prime }=4\pi a\hbar
^{2}/m$, with $a$ the $s$-wave scattering length. Eq.~(\ref{1}) is obtained
under the assumption that the interaction potential can be represented as a
zero-range pseudo-potential of strength $g^{\prime }$. The equation for the
condensate wave-function $\Psi \left( t,\vec{r}\right) =\left\langle \hat{%
\Phi}\left( t,\vec{r}\right) \right\rangle $ is obtained by taking the
average of Eq.~(\ref{1}) with respect to a broken symmetry non-equilibrium
ensemble in which the quantum field operator takes a non-zero expectation
value. Then for the exact equation of motion of $\Psi \left( t,\vec{r}\right) $
we find
\begin{equation}
i\hbar \frac{\partial \Psi \left( t,\vec{r}\right) }{\partial t}=\left[ -%
\frac{\hbar ^{2}}{2m}\Delta +mV_{grav}\left( \vec{r}\right) \right] \Psi
\left( t,\vec{r}\right) +g^{\prime }\left\langle \hat{\Phi}^{+}\left( t,\vec{%
r}\right) \hat{\Phi}\left( t,\vec{r}\right) \hat{\Phi}\left( t,\vec{r}%
\right) \right\rangle .  \label{2a}
\end{equation}

By introducing the non-condensate field operator $\tilde{\psi}\left( t,\vec{r%
}\right) $ we separate out the condensate component of the quantum field
operator, so that \cite{Za99,Prouk, Gr}
\begin{equation}
\hat{\Phi}\left( t,\vec{r}\right) =\Psi \left( t,\vec{r}\right) +\tilde{\psi}%
\left( t,\vec{r}\right) ,  \label{3}
\end{equation}
where the average value of $\tilde{\psi}\left( t,\vec{r}\right) $ is zero, $%
\left\langle \tilde{\psi}\left( t,\vec{r}\right) \right\rangle =0$. By
taking into account the relations
\begin{equation}
\hat{\Phi}^{+}\left( t,\vec{r}\right) \hat{\Phi}\left( t,\vec{r}\right) \hat{%
\Phi}\left( t,\vec{r}\right) =\left| \Psi \right| ^{2}\Psi +2\left| \Psi
\right| ^{2}\tilde{\psi}+\Psi ^{2}\tilde{\psi}^{+}+\Psi ^{\ast }\tilde{\psi}%
\tilde{\psi}+2\Psi \tilde{\psi}^{+}\tilde{\psi}+\tilde{\psi}^{+}\tilde{\psi}%
\tilde{\psi},
\end{equation}
we obtain
\begin{equation}
\left\langle \hat{\Phi}^{+}\left( t,\vec{r}\right) \hat{\Phi}\left( t,\vec{r}%
\right) \hat{\Phi}\left( t,\vec{r}\right) \right\rangle =\frac{1}{m}\rho
_{c}\Psi +2\frac{1}{m}\tilde{\rho}\Psi +\rho _{\tilde{m}}\Psi ^{\ast
}+\left\langle \tilde{\psi}^{+}\tilde{\psi}\tilde{\psi}\right\rangle ,
\label{5}
\end{equation}
where we have introduced the local condensate mass density $\rho _{c}\left(
t,\vec{r}\right) =mn_{c}\left( t,\vec{r}\right) =m\left| \Psi \left( t,\vec{r%
}\right) \right| ^{2}$, the non-condensate mass density $\tilde{\rho}\left(
t,\vec{r}\right) =m\tilde{n}\left( t,\vec{r}\right) =m\left\langle \tilde{%
\psi}^{+}\left( t,\vec{r}\right) \tilde{\psi}\left( t,\vec{r}\right)
\right\rangle $, and the off-diagonal (anomalous) mass density $\rho _{%
\tilde{m}}=m\tilde{m}\left( t,\vec{r}\right) =m\left\langle \tilde{\psi}%
\left( t,\vec{r}\right) \tilde{\psi}\left( t,\vec{r}\right)
\right\rangle $, respectively. With the use of Eq.~(\ref{5}) in Eq.~(\ref{2a}%
) we obtain the equation of motion for $\Psi $ as \cite{Za99,Prouk, Gr}
\begin{eqnarray}
i\hbar \frac{\partial \Psi \left( t,\vec{r}\right) }{\partial t}&=&\left[ -%
\frac{\hbar ^{2}}{2m}\Delta +mV_{grav}\left( \vec{r}\right) +g\rho _{c}\left(
t,\vec{r}\right) +2g\tilde{\rho}\left( t,\vec{r}\right) \right] \Psi \left(
t,\vec{r}\right) +  \notag \\
&&g\rho _{\tilde{m}}\Psi ^{\ast }+g^{\prime }\left\langle \tilde{\psi}%
^{+}\left( t,\vec{r}\right) \tilde{\psi}\left( t,\vec{r}\right) \tilde{\psi}%
\left( t,\vec{r}\right) \right\rangle ,  \label{6}
\end{eqnarray}
where we have denoted $g=4\pi a\hbar ^{2}/m^{2}$. Besides the condensate
density $\rho _{c}\left( t,\vec{r}\right) $ the generalized Gross-Pitaevskii
Eq.~(\ref{6}) also contains the non-condensate density $\tilde{\rho}\left( t,%
\vec{r}\right) $, the off-diagonal non-condensate density $\rho _{\tilde{m}}$%
, and the three-field correlation function $\left\langle \tilde{\psi}^{+}%
\tilde{\psi}\tilde{\psi}\right\rangle $, both of which have nonzero
expectation values due to the assumed Bose broken symmetry.

The generalized Gross-Pitaevski equation can be transformed to a
hydrodynamic form by introducing the Madelung representation of the wave
function as $\Psi \left( t,\vec{r}\right) =\sqrt{\rho _{c}}\exp \left[
\left( i/\hbar \right) S\left( t,\vec{r}\right) \right] $. Then it follows
that Eq.~(\ref{6}) is equivalent to the system \cite{Za99, Gr}
\begin{equation}
\frac{\partial \rho _{c}}{\partial t}+\nabla \cdot \left( \rho _{c}\vec{v}%
_{c}\right) =2\frac{g}{\hbar }\mathrm{Im}\left[ \left( \Psi ^{\ast }\right) ^{2}\rho _{%
\tilde{m}}+m\Psi ^{\ast }\left\langle \tilde{\psi}^{+}\tilde{\psi}\tilde{\psi%
}\right\rangle \right] =-\Gamma _{12},
\end{equation}
\begin{equation}
\frac{\partial S}{\partial t}=-\left( \mu _{c}+\frac{1}{2}m\vec{v}%
_{c}^{2}\right) ,  \label{7}
\end{equation}
where the local velocity of the condensate is given by $\vec{v}_{c}\left( t,\vec{r}%
\right) =\left( \hbar /m\right) \nabla S$ and  we have introduced a new function
$\Gamma _{12}$, which plays the role of the local source term in the
condensate continuity equation. The chemical potential
of the condensate is defined as
\begin{equation}
\mu _{c}=-\frac{\hbar ^{2}}{2m}\frac{\Delta \sqrt{\rho _{c}}}{\sqrt{\rho _{c}%
}}+mV_{grav}\left( \vec{r}\right) +g\rho _{c}\left( t,\vec{r}\right) +2g\tilde{%
\rho}\left( t,\vec{r}\right) +\mathrm{Re}R\left( t,\vec{r}\right) ,
\end{equation}
where the dissipation term $R(t,\vec{r})$ is given by
\begin{equation}
R\left( t,\vec{r}\right) =\frac{g}{ \rho _{c}}\left[ \left( \Psi ^{\ast
}\right) ^{2}\rho _{\tilde{m}}+\Psi ^{\ast }\left\langle \tilde{\psi}^{+}%
\tilde{\psi}\tilde{\psi}\right\rangle \right] .
\end{equation}

Eq.~(\ref{7}) can be reformulated as the Euler equation of fluid dynamics for the condensate,
\begin{equation}
m\frac{d\vec{v}_{c}}{dt}=m\left[ \frac{\partial \vec{v}_{c}}{\partial t}%
+\left( \vec{v}_{c}\cdot \nabla \right) \vec{v}_{c}\right] =-\nabla \mu _{c}.
\end{equation}

There are a number of approximations that can be introduced in Eq.~(\ref{6}%
), and which have been investigated in detail in the literature. If all the
particles are in the condensate, $\tilde{\rho}\left( t,\vec{r}\right) =0$,
and the anomalous correlations $\rho _{\tilde{m}}$ and $\left\langle \tilde{%
\psi}^{+}\tilde{\psi}\tilde{\psi}\right\rangle $ are absent, Eq.~(\ref{6})
reduces to the well-known Gross-Pitaevskii equation \cite{Pe}.

The Gross-Pitaevskii
equation gives a very good description of the condensate if $T<<T_{BEC}$. In
fact the Gross-Pitaevskii equation gives a good quantitative description of
the condensate for $T\preceq 0.5T_{BEC}$, a temperature range for which the
interaction induced depletion of the condensate is estimated to be of the
order of a few percent \cite{GP,Gr}. The physical reason for this possibility is that
in the limit where the modes of the system
are highly occupied ($N_k >>1$), the classical fluctuations
of the field overwhelm the quantum fluctuations. Therefore
these modes may  be represented by a coherent
wave function. This is analogous to the situation
in laser physics, where the highly occupied laser modes
can be well described by classical equations.

In the Hartree-Fock-Bogoliubov (HFB) approximation one ignores the
three-field correlation function $\left\langle \tilde{\psi}^{+}\tilde{\psi}%
\tilde{\psi}\right\rangle $, and the $\rho _{c}\left( t,\vec{r}\right) $, $%
\tilde{\rho}\left( t,\vec{r}\right) $, and $\rho _{\tilde{m}}$ fluctuations
are calculated self-consistently \cite{HFB}. The dynamic Popov approximation consists
in ignoring both $\left\langle \tilde{\psi}^{+}\tilde{\psi}\tilde{\psi}%
\right\rangle $ and $\rho _{\tilde{m}}$ \cite{Pop}. In the static Popov approximation
the fluctuations of the density of the thermal cloud are ignored by assuming
that the non-condensate is always in static thermal equilibrium, so that $%
\tilde{\rho}\left( t,\vec{r}\right) \approx \tilde{\rho}\left( \vec{r}%
\right) $ \cite{Da99}.

In the static case the anomalous density $\rho _{\tilde{m}}$ can be obtained
as $\rho _{\tilde{m}}=-g\rho _{c}\left( 1/V\right) \int \left[ \left(
1+2f_{p}^{0}\right) /2E_{p}\right] d\vec{p}/\left( 2\pi \hbar \right) ^{3}$,
where $E_{p}$ is the local excitation energy (per unit mass), $V$ is the
volume of the system, and $f_{p}^{0}$ is the equilibrium Bose-Einstein
distribution function \cite{Za99}. The anomalous density, as well as the three-field
correlation function, also depends on the volume of the system. Since the
integral takes values of the order of unity, one can approximate the
anomalous density as  $\rho _{\tilde{m}}\sim f\left(kT/E_p\right)g\rho _{c}/\left(VE_p\right)$,
where $f\left(kT/E_p\right)$ is a function of the temperature with numerical values of the
order of unity or less. Therefore the contribution of the anomalous density
term in the generalized Gross-Pitaevskii equation Eq.~(\ref{6}) is of second
order in the coupling constant $g<1$, and its contribution to the total
density can generally be neglected with respect to the condensate density $%
\rho _{c}$. In   the limit of high temperatures  $\rho _{\tilde{m}}/\rho
_{c}\sim \left( g/VE_{p}\right) \left( kT/E_{p}\right) $, while in the
limit of $T\rightarrow 0$, $\rho _{\tilde{m}}/\rho _{c}\sim \left(
g/VE_{p}\right) $.  In a low temperature system and in laboratory conditions the contribution of the
anomalous density to the total density is of the order of a few percents.
However, once $T$ approaches the critical temperature $T_{c}$, the anomalous
density can play a significant role in the dynamics of the condensate. The
three-field correlation function has the same order of magnitude as the
anomalous density, thus also giving a second order correction in the
generalized Gross-Pitaevskii equation. In the limit of $V\rightarrow \infty $, a situation that represent an excellent approximation in the case of dark
matter halos, with radii of the order of a few tenth of kiloparsecs, both
the anomalous density and the three-field correlation function tend
rigourously to zero.

\section{Finite temperature Bose-Einstein condensed dark matter density
profiles}\label{3_s}

In the following we will restrict our analysis to the range of finite
temperatures where the dominant thermal excitations can be approximated as
high energy non-condensed particles moving in a self-consistent Hartree-Fock
mean field, with local energy \cite{Za99, Prouk, Gr}
\begin{equation}
\bar{\varepsilon}_{p}\left( t,\vec{r}\right) =\frac{\vec{p}^{2}}{2m}%
+mV_{grav}\left( \vec{r}\right) +2g\left[ \rho _{c}\left( t,\vec{r}\right) +%
\tilde{\rho}\left( t,\vec{r}\right) \right] =\frac{\vec{p}^{2}}{2m}%
+U_{eff}\left( t,\vec{r}\right) ,
\end{equation}
where $U_{eff}\left( t,\vec{r}\right) =mV_{grav}\left( \vec{r}\right) +2g\left[
\rho _{c}\left( t,\vec{r}\right) +\tilde{\rho}\left( t,\vec{r}\right) %
\right] $. Hence in the present approximation we neglect the effects of
the mean field associated with the anomalous density $\rho _{\tilde{m}}$ and
the three-field correlation function $\left\langle \tilde{\psi}^{+}\tilde{%
\psi}\tilde{\psi}\right\rangle $.

Moreover, we consider the case in which
the thermal cloud and the condensate are in static equilibrium. Therefore
there are no stationary equilibrium currents, $\vec{v}_{c}=0$, and the
source and dissipation terms in the hydrodynamic equations of motion vanish,
$\Gamma _{12}=0$, $R=0$.

In order to describe the gravitational properties of the Bose-Einstein condensed dark matter halo we consider the gravitational interaction within a mean field approximation, by introducing the self-consistent gravitational potential as $V_{grav}\left( t,\vec{r}\right) =\int G\left( \vec{r}-\vec{r}^{\prime
}\right) \left\vert \Psi \left( t,\vec{r}\right) \right\vert ^{2}d\vec{r}$, where $G\left( \vec{r}-\vec{r}^{\prime }\right) =-Gm^{2}/\left\vert \vec{r}-\vec{r}%
^{\prime }\right\vert $ is the usual gravitational interaction potential. Therefore the Gross-Pitaevskii equation describes a self-gravitating Bose-Einstein condensate with short range interactions. The gravitational potential $V_{grav}\left( t,\vec{r}\right)$ is determined self-consistently  by the Poisson equation, given by
\begin{equation}
\Delta V_{grav}=4\pi G\left[ \rho _{c}\left( \vec{r}\right) +\tilde{\rho}%
\left( \vec{r}\right) \right] .  \label{8}
\end{equation}

Under these assumption, and by representing the equilibrium condensate wave
function as $\Psi \left( t,\vec{r}\right) =\Psi _{0}\left( \vec{r}\right)
\exp \left[ -(i/\hbar )\mu _{c}t\right] $, $\mu _{c}=$constant, it follows
that the stationary static condensate wave function $\Psi _{0}\left( \vec{r}%
\right) $ satisfies the equation
\begin{equation}\label{chempot}
\left[ -\frac{\hbar ^{2}}{2m}\Delta +mV_{grav}\left( \vec{r}\right) +g\rho
_{c}\left( \vec{r}\right) +2g\tilde{\rho}\left( \vec{r}\right) \right]
\Psi _{0}\left( \vec{r}\right) =\mu _{c}\Psi _{0}\left( \vec{r}\right) .
\end{equation}

In the static case the energy of the thermal excitations is given by
\begin{equation}
\bar{\varepsilon}_{p}\left( \vec{r}\right) =\frac{\vec{p}^{2}}{2m}%
+mV_{grav}\left( \vec{r}\right) +2g\left[ \rho _{c}\left( \vec{r}\right) +%
\tilde{\rho}\left( \vec{r}\right) \right] =\frac{\vec{p}^{2}}{2m}%
+U_{eff}\left( \vec{r}\right) ,
\end{equation}
where the gravitational potential is modified by the Hartree-Fock mean field, $%
U_{eff}\left( \vec{r}\right) =mV_{grav}\left( \vec{r}\right) +2g\left[ \rho
_{c}\left( \vec{r}\right) +\tilde{\rho}\left( \vec{r}\right) \right] $.

The collision between particles in the thermal cloud forces a
non-equilibrium distribution to evolve to the static Bose-Einstein
distribution $f^{0}\left( \vec{r},\vec{p}\right) $. Hence the particles in
the thermal cloud are in thermodynamic equilibrium among themselves. By
using a single-particle representation spectrum the equilibrium distribution
of the thermal cloud is given by
\begin{equation}
f^{0}\left( \vec{r},\vec{p}\right) =\left[ e^{\beta \bar{\varepsilon}%
_{p}\left( \vec{r}\right) -\tilde{\mu}}-1\right] ^{-1},
\end{equation}
where $\beta =1/k_{B}T$, with $k_B$ Boltzmann's constant, and $\tilde{\mu}$ is the chemical potential of the
thermal cloud. In order to determine $\tilde{\mu}$ we assume that the condensate and the thermal cloud components are in local diffusive equilibrium with respect to each other. The requirement of a
static diffusive equilibrium between the cloud and the condensate imposes
the condition \cite{Za99,Gr}
\be
\mu _{c}=\tilde{\mu}.
\ee
Therefore the chemical potential
of the condensate also determines the static equilibrium distribution of the
particles in the cloud.

As a last step in our analysis we introduce the Thomas-Fermi approximation
for the condensate wave function. In the Thomas-Fermi approximation, the
kinetic energy term $-\left( \hbar ^{2}/2m\right) \Delta $ of the condensate
particles is neglected.
In order to analyze the validity of the Thomas-Fermi approximation we consider a system of $N$ bosons, all in the same state, occupying a volume $V$, with spatial extent $R$, and confined by a gravitational potential. The  mass of the system is $M$, while for the mass and the scattering length of the dark matter particles we adopt the values $m=3.19\times 10^{-37}$ g and $a=1.78\times 10^{-19}$ cm, respectively (see Section \ref{4}). The total energy $E$ of the system can be written as  $E=E_{kin}+E_{int}+E_{grav}$, where $E_{kin}$, $E_{int}$ and $E_{grav}$ are the kinetic energy, interaction energy, and gravitational energy, respectively. The kinetic energy per particle is $\hbar ^2/2mR^2$ \cite{Pe}, and therefore the total kinetic energy of the system is given by $E_{kin}=N\hbar ^2/2mR^2$. The interaction energy can be obtained as $E_{int}=(1/2) \left(N^2/V\right)mg$ \cite{Pe}, while the gravitational potential energy is $E_{grav}=GM^2/R$. Therefore the total energy of the system can be written as
\be
E=N\frac{\hbar ^2}{2mR^2}+\frac{3}{2}N^2\frac{\hbar ^2a}{mR^3}+\frac{GM^2}{R}.
\ee

The interaction energy is much larger than the kinetic energy, $E_{int}>>E_{kin}$ if the condition
\be
\frac{3Na}{R}>>1
\ee
is satisfied. By assuming that the mass of the condensate is $M=10^{10}M_{\odot}=2\times 10^{43}$ g, the dark matter particle number is of the order $N=M/m=6.2\times 10^{79}$ particles. For a condensate with radius  $R=10$ kpc = $3\times 10^{22}$ cm, the quantity $3Na/R=1.11\times 10^{39}$, which is obviously much greater than one, $3Na/R>>1$. Therefore the contribution of the kinetic energy of the particles can be neglected with respect to the interaction energy of the bosons. Hence in the case of a large particle number, the Thomas-Fermi approximation gives an excellent description of the condensate, at such a precision level that the present description of the dark matter halos can be considered {\it exact}. The Thomas-Fermi approximation breaks down at length scales of the order of $R\approx \sqrt{m/4\pi \rho a}$, where $\rho $ is the mean dark matter density. For $\rho =10^{-24}\;{\rm g/cm^3}$, we obtain $R\approx 377.65$ cm, a length scale that is irrelevant from an astronomical point of view. The gravitational energy of the dark matter halo has the value $E_{grav}=8.89\times 10^{56}$ ergs, why for the considered parameters $E_{int}=1.35\times 10^{56}$ ergs, which is of the same order of magnitude as the gravitational energy. However, $E_{grav}>E_{int}$, thus showing that the gravitational energy is indeed a trapping energy for the dark matter halo.

Hence we can neglect, with a very good approximation, the kinetic energy term in Eq.~(\ref{chempot}), and thus the chemical potential of the condensate (and of the thermal cloud) is given by
\begin{equation}  \label{8b}
\mu _{c}=\tilde{\mu}=mV_{grav}\left( \vec{r}\right) +g\left[ \rho _{c}\left(
\vec{r}\right) +2\tilde{\rho}\left( \vec{r}\right) \right] .
\end{equation}

The equilibrium non-condensate density is obtained by integrating the
equilibrium Bose - Einstein distribution over the momentum. Thus we obtain \cite{Prouk, Gr}
\begin{equation}
\tilde{\rho}\left( \vec{r}\right) =\frac{m}{\left(2\pi \hbar \right)^{3}}\int d^{3}\vec{p}%
f^{0}\left( \vec{r},\vec{p}\right) =\frac{m}{\lambda _T^{3}}g_{3/2}\left[
z\left( \vec{r}\right) \right] ,
\end{equation}
where $\lambda _T=$ $\sqrt{2\pi \hbar ^{2}\beta /m}$ is the de Broglie thermal
wavelength, $g_{3/2}(z)$ is a Bose-Einstein function, and the fugacity $%
z\left( \vec{r}\right) $ is given by
\begin{equation}
z\left( \vec{r}\right) =e^{\beta \left[ \tilde{\mu}-U_{eff}\left( \vec{r}%
\right) \right] }=e^{-\beta g\rho _{c}\left( \vec{r}\right) }.
\end{equation}

By applying the operator $\Delta $ on both sides of Eq.~(\ref{8b}), representing the conservation of the energy, and by taking into account that $\mu _{c}$ and $\tilde{\mu}$ are constants,  we obtain immediately
\begin{equation}
g\Delta \left[ \rho _{c}\left( \vec{r}\right) +2\tilde{\rho}\left( \vec{r}%
\right) \right] =-m\Delta V_{grav}=-4\pi Gm\left[ \rho _{c}\left(\vec{r}%
\right) +\tilde{\rho}\left(\vec{r}\right) \right] .
\end{equation}
Therefore the static condensate density profile at non-zero temperature is
described by the following equation
\begin{equation}
\Delta \left\{ \rho _{c}\left( \vec{r}\right) +2\frac{m}{\lambda _T^{3}}g_{3/2}%
\left[ e^{-\beta g\rho _{c}\left( \vec{r}\right) }\right] \right\} =-\frac{%
4\pi Gm}{g}\left\{ \rho _{c}\left( \vec{r}\right) +\frac{m}{\lambda _T^{3}}%
g_{3/2}\left[ e^{-\beta g\rho _{c}\left( \vec{r}\right) }\right] \right\} .
\label{10a}
\end{equation}

In order to give an approximate representation of the Bose-Einstein functions $g_{3/2}%
\left\{ \exp\left[-\beta g\rho _{c}\left( \vec{r}\right)\right]\right\}$ appearing in Eq.~(\ref{10a}) we estimate first
the numerical values of the quantity $-\beta g\rho _{c}\left( \vec{r}\right)$. 
As a first numerical approximation we take for the density profile its maximum value at the center, so that  
$\rho _{c}\left( \vec{r}\right)\approx \rho _{c}\left( 0\right)$. 
For realistic galactic dark matter halos the central density of the dark matter is in the range of 
$\rho _{c}\left( 0\right)\approx 10^{-24}-10^{-26}$ g/cm$^3$ \cite{OvWe04}.   
By adopting for the mass and for the scattering length of the dark matter particles the values 
$m=3.19\times 10^{-37}$ g and $a=1.78\times 10^{-19}$ cm, respectively,  
we obtain an upper bound for $\beta g\rho _{c}\left( 0\right)$ as $\left\{\max \beta g\rho _{c}\left( 0\right)\right\}\in \left( 1.7695\times 10^{-7}/T,1.7695\times 10^{-9}/T\right)$.
Hence in any physically realistic dark matter scenario in which the temperature of the dark matter halo 
is higher than the temperature of the Cosmic Microwave Background Radiation,  $T>2.73$ K, the quantity 
$\beta g\rho _{c}\left( \vec{r}\right)<<1$, for all $\vec{r}$ and $T$. As a next step we  obtain a linear approximation of the function 
$g_{3/2}%
\left\{ \exp\left[-\beta g\rho _{c}\left( \vec{r}\right)\right]\right\}$. In order to estimate $g_{3/2}(x)$ we make 
{\it two linear approximations simultaneously}, namely, 
\begin{equation}
g_{3/2}(x)\approx g_{3/2}(1)+g_{1/2}\left(1\right)\times \Delta x,
\end{equation}
with $x=
 \exp\left[-\beta g\rho _{c}\left( \vec{r}\right)\right]$, and 
\begin{equation}
\Delta x=  e^{-\beta g\rho _{c}\left( \vec{r}\right) }-1\approx -\beta g\rho _{c}\left( \vec{r}\right).
\end{equation}
 
Therefore in the first order of approximation the Bose-Einstein function $g_{3/2}%
\left\{ \exp\left[-\beta g\rho _{c}\left( \vec{r}\right)\right]\right\}$ can be obtained as
\begin{equation}
g_{3/2}\left[ e^{-\beta g\rho _{c}\left( \vec{r}\right) }\right] \approx
\zeta \left( \frac{3}{2}\right) -\zeta \left( \frac{1}{2}\right) \beta g\rho
_{c}\left( \vec{r}\right) +...,
\end{equation}%
where $\zeta \left( 3/2\right)=g_{3/2}(1) =2.612$ and $\zeta \left( 1/2\right) =g_{1/2}(1)=-1.460$
are Riemann zeta functions. Hence the density profile of the condensate component of the finite
temperature dark matter halo can be obtained by solving the second order differential equation given by
\begin{equation}
\Delta \rho _{c}\left( \vec{r}\right) =-\frac{4\pi Gm}{g}\frac{1-\zeta
\left( 1/2\right) mg\beta /\lambda _T^{3}}{1-2\zeta \left( 1/2\right) mg\beta
/\lambda _T^{3}}\rho _{c}\left( \vec{r}\right) -\frac{4\pi \zeta (3/2)Gm^{2}}{%
g\lambda _T^{3}\left[ 1-2\zeta \left( 1/2\right) mg\beta /\lambda _T^{3}\right] }%
.
\end{equation}

 By representing the condensate density as
\begin{equation}
\rho _{c}\left( \vec{r}\right) =\rho _{c}^{(0)}+\rho _{c}^{(1)}\left( \vec{r}%
\right) ,
\end{equation}%
where
\begin{equation}
\rho _{c}^{(0)}=-\frac{\zeta (3/2)m}{\lambda _T^{3}\left[ 1-\zeta \left(
1/2\right) mg\beta /\lambda _T^{3}\right] },
\end{equation}
it follows that the function $\rho _{c}^{(1)}\left( \vec{r}\right) $
satisfies the equation
\begin{equation}
\Delta \rho _{c}^{(1)}\left( \vec{r}\right) =-K^{2}\rho _{c}^{(1)}\left(
\vec{r}\right) ,  \label{11a}
\end{equation}%
where we have denoted
\begin{equation}
K^{2}=\frac{Gm^{3}}{a\hbar ^{2}}\frac{1-\zeta \left( 1/2\right) mg\beta
/\lambda _T^{3}}{1-2\zeta \left( 1/2\right) mg\beta /\lambda _T^{3}}.
\end{equation}

In static spherical symmetry the general non-singular solution of Eq.~(\ref%
{11a}) is given by
\begin{equation}
\rho _{c}^{(1)}\left( r\right) =A\frac{\sin Kr}{Kr},
\end{equation}%
where $A$ is an arbitrary integration constant. The density of the
non-condensate component of the dark matter can be written as
\begin{equation}
\tilde{\rho}\left( r\right) \approx \frac{m}{\lambda _T^{3}}\zeta \left( \frac{%
3}{2}\right) -\zeta \left( \frac{1}{2}\right) \frac{mg\beta }{\lambda _T^{3}}%
\rho _{c}\left( r\right) ,
\end{equation}%
while the total density of the finite temperature dark matter halo is given
by
\begin{equation}
\rho _{tot}(r)=\rho _{c}(r)+\tilde{\rho}\left( r\right) =\frac{m}{\lambda _T
^{3}}\zeta \left( \frac{3}{2}\right) +\rho _{c}^{(0)}+A\left[ 1-\zeta \left(
\frac{1}{2}\right) \frac{mg\beta }{\lambda _T^{3}}\right] \frac{\sin Kr}{Kr}.
\end{equation}

At the center of the halo the density of the dark matter is $\rho
_{tot}(0)=\rho _{tot}^{(0)}$, which allows us to determine the integration
constant $A$ as
\begin{equation}
A\left[ 1-\zeta \left(
\frac{1}{2}\right) \frac{mg\beta }{\lambda _T^{3}}\right]=\rho _{tot}^{(0)}+\frac{\zeta (3/2)m}{\lambda _T^{3}\left[ 1-2\zeta \left(
1/2\right) mg\beta /\lambda _T^{3}\right] }-\frac{m}{\lambda _T^{3}}\zeta \left( \frac{3}{2}%
\right) .
\end{equation}

The total density profile of the dark matter halo can be represented in the form
\begin{equation}
\rho _{tot}(r)=\rho _{tot}^{(0)}\frac{\sin Kr}{Kr}+\left[ \frac{m}{\lambda _T
^{3}}\zeta \left( \frac{3}{2}\right) +\rho _{c}^{(0)}\right] \left( 1-\frac{%
\sin Kr}{Kr}\right) .
\end{equation}

The previous results can be written in a more transparent form if we introduce the condensation temperature $T_{BEC}$, given by $T_{BEC}=2\pi \hbar ^2\rho _{BEC}^{2/3}/\zeta ^{2/3}\left(3/2\right)m^{5/3}k_B$, where $\rho _{BEC}$ is the density of the dark matter at the condensation moment. Then we obtain immediately
\be
\frac{m}{\lambda _T^3}=\frac{\rho _{BEC}}{\zeta (3/2)}\left(\frac{T}{T_{BEC}}\right)^{3/2}, g\beta =2\zeta ^{2/3}(3/2)\frac{a}{m^{1/3}\rho _{BEC}^{2/3}}\left(\frac{T}{T_{BEC}}\right)^{-1},
\ee
and
\be
\frac{mg\beta }{\lambda _T^3}=\frac{2}{\zeta ^{1/3}(3/2)}\frac{a}{m^{1/3}}\rho _{BEC}^{1/3}\left(\frac{T}{T_{BEC}}\right)^{1/2},
\ee
respectively. Hence we obtain
\be
K^{2}(T)=\frac{Gm^{3}}{a\hbar ^{2}}\frac{1-\alpha \left(a/m^{1/3}\right)\rho _{BEC}^{1/3}\left(T/T_{BEC}\right)^{1/2}}{1-2\alpha \left(a/m^{1/3}\right)\rho _{BEC}^{1/3}\left(T/T_{BEC}\right)^{1/2}},
\ee
where we have denoted $\alpha =2\zeta (1/2)/\zeta ^{1/3}(3/2)<0$. By using a series expansion for $\zeta \left( 1/2\right) mg\beta /\lambda _T
^{3}<<1$, $K^{2}(T)$ can be approximated as
\begin{equation}
K^{2}(T)\approx \frac{Gm^{3}}{a\hbar ^{2}}\left[ 1+\zeta \left( \frac{1}{2}\right) mg%
\frac{\beta }{\lambda _T^{3}}\right] ,
\end{equation}
or
\be
K^{2}(T)\approx \frac{Gm^{3}}{a\hbar ^{2}}\left[ 1+\alpha \frac{a}{m^{1/3}}\rho _{BEC}^{1/3}\left(\frac{T}{T_{BEC}}\right)^{1/2}\right] .
\ee
For the total density profile of the dark matter halo we obtain
\bea\label{prof}
\rho _{tot}(r)&=&\left[ \rho _{tot}^{(0)}+\frac{\alpha \left(
a/m^{1/3}\right) \rho _{BEC}^{4/3}\left( T/T_{BEC}\right) ^{2}}{1-\alpha
\left( a/m^{1/3}\right) \rho _{BEC}^{1/3}\left( T/T_{BEC}\right) ^{1/2}}%
\right]\frac{\sin \left[K(T)r\right]}{K(T)r}-\nonumber\\
&&\frac{\alpha \left(
a/m^{1/3}\right) \rho _{BEC}^{4/3}\left( T/T_{BEC}\right) ^{2}}{1-\alpha
\left( a/m^{1/3}\right) \rho _{BEC}^{1/3}\left( T/T_{BEC}\right) ^{1/2}},
\eea
or, equivalently,
\be
\rho _{tot}(r)=\left[ \rho _{tot}^{(0)}+\rho _T(T)%
\right]\frac{\sin \left[K(T)r\right]}{K(T)r}-\rho _T(T),
\ee
where we have denoted
\be
\rho_T(T)=\frac{\alpha \left(
a/m^{1/3}\right) \rho _{BEC}^{4/3}\left( T/T_{BEC}\right) ^{2}}{1-\alpha
\left( a/m^{1/3}\right) \rho _{BEC}^{1/3}\left( T/T_{BEC}\right) ^{1/2}}<0.
\ee

The mass profile $M_{tot}(r)=4\pi \int _0^r{\rho _{tot}(r)r^2dr}$ of the dark matter halo is given by
\begin{eqnarray}
M_{tot}\left( r\right)  &=&-\frac{4\pi \alpha \left( a/m^{1/3}\right) \rho
_{BEC}^{4/3}\left( T/T_{BEC}\right) ^{2}}{3\left[ 1-\alpha \left(
a/m^{1/3}\right) \rho _{BEC}^{1/3}\left( T/T_{BEC}\right) ^{1/2}\right] }%
r^{3}+ \nonumber\\
&&\frac{1}{K^{3}(T)}\left[ \rho _{tot}^{(0)}+\frac{\alpha \left(
a/m^{1/3}\right) \rho _{BEC}^{4/3}\left( T/T_{BEC}\right) ^{2}}{1-\alpha
\left( a/m^{1/3}\right) \rho _{BEC}^{1/3}\left( T/T_{BEC}\right) ^{1/2}}%
\right]\times \nonumber\\
&& \left\{ \sin \left[K(T)r\right]-K(T)r\cos \left[K(T)r\right]\right\} ,
\end{eqnarray}
or, equivalently,
\be
M_{tot}\left( r\right)=-\frac{4\pi }{3}\rho _T (T)r^3+\frac{1}{K^3(T)}\left[ \rho _{tot}^{(0)}+\rho _T(T)%
\right]\left\{ \sin \left[K(T)r\right]-K(T)r\cos \left[K(T)r\right]\right\}.
\ee

 The radius $R$ of the dark matter distribution is determined by the condition $\rho _{tot}(R)=0$. In the zero-temperature limit $T\rightarrow 0$, the radius $R_0$ of the halo is given by $R_0=\pi/K(0)$. By representing $K(T)R$ as $K(T)R=\pi+\phi _T\pi $, we obtain first $\sin[K(T)R]=-\sin \left(\phi _T\pi \right)\approx -\phi _T\pi $. Hence for $\phi _T$ we obtain
 \be
 \phi _T(T)=-\frac{\rho _T(T)}{\rho _{tot}^{(0)}+2\rho _T(T)}.
 \ee
 Therefore the radius of the dark matter halo at a finite temperature $T$ is given by
 \be
 R(T)\approx \frac{\pi }{K(T)}\left[1-\frac{\rho _T(T)}{\rho _{tot}^{(0)}+2\rho _T(T)}\right].
 \ee

\begin{figure}[!ht]
\includegraphics[width=0.98\linewidth]{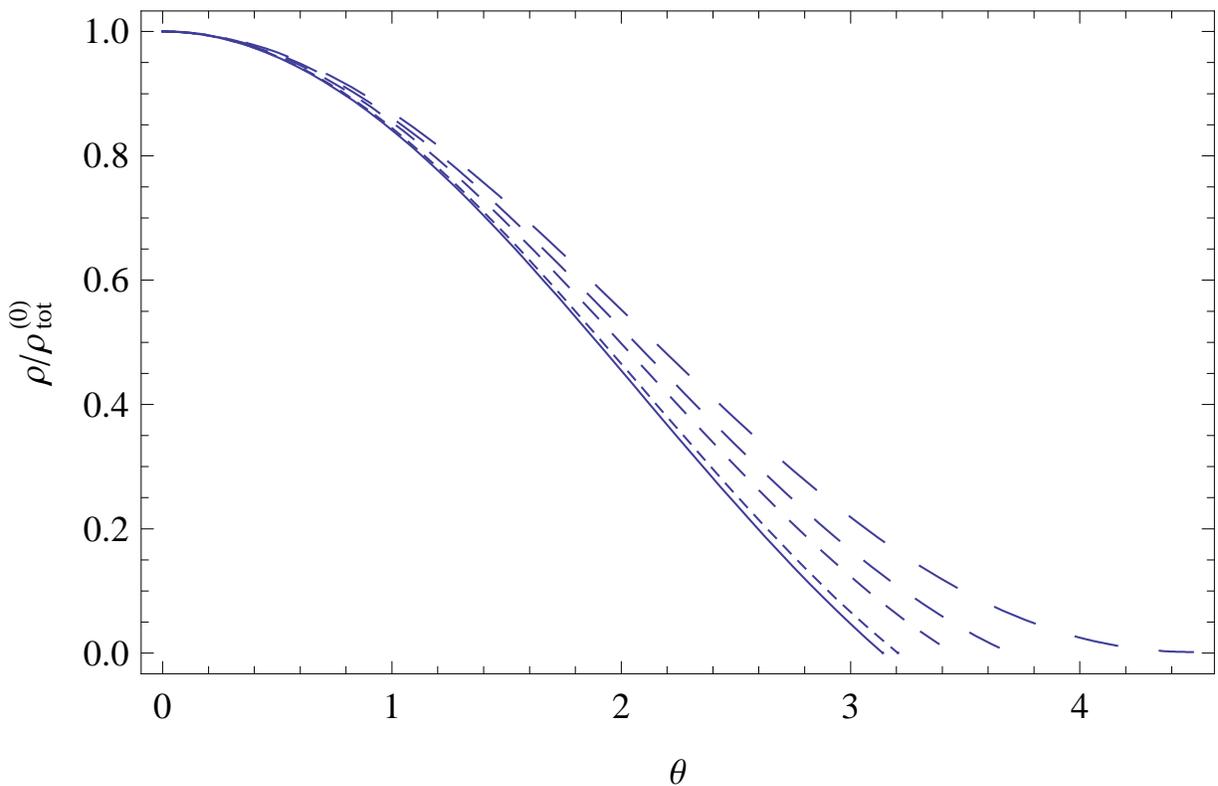}
\caption{Radial coordinate dependence of the total density profile of the finite temperature dark matter halo, for different values of the temperature: $T=0$ (solid curve), $T/T_{BEC}=0.2$ (dotted curve), $T/T_{BEC}=0.4$ (dashed curve), $T/T_{BEC}=0.5$ (long dashed curve), and $T/T_{BEC}=0.6$ (ultra-long dashed curve). For all the curves $\alpha \left(a/m^{1/3}\right)=-5000\;{\rm cm/g^{1/3}}$. The Bose-Einstein transition density is taken as $\rho _{BEC}=10^{-21}\;{\rm g/cm^3}$, while the central density of the dark matter halo is $\rho _{tot}^{(0)}=10^{-24}\;{\rm g/cm^3}$.  The dimensionless radial coordinate $\theta $ is defined as $\theta =K(T)r$.  }
\label{FIG1}
\end{figure}

\begin{figure}[!ht]
\includegraphics[width=0.98\linewidth]{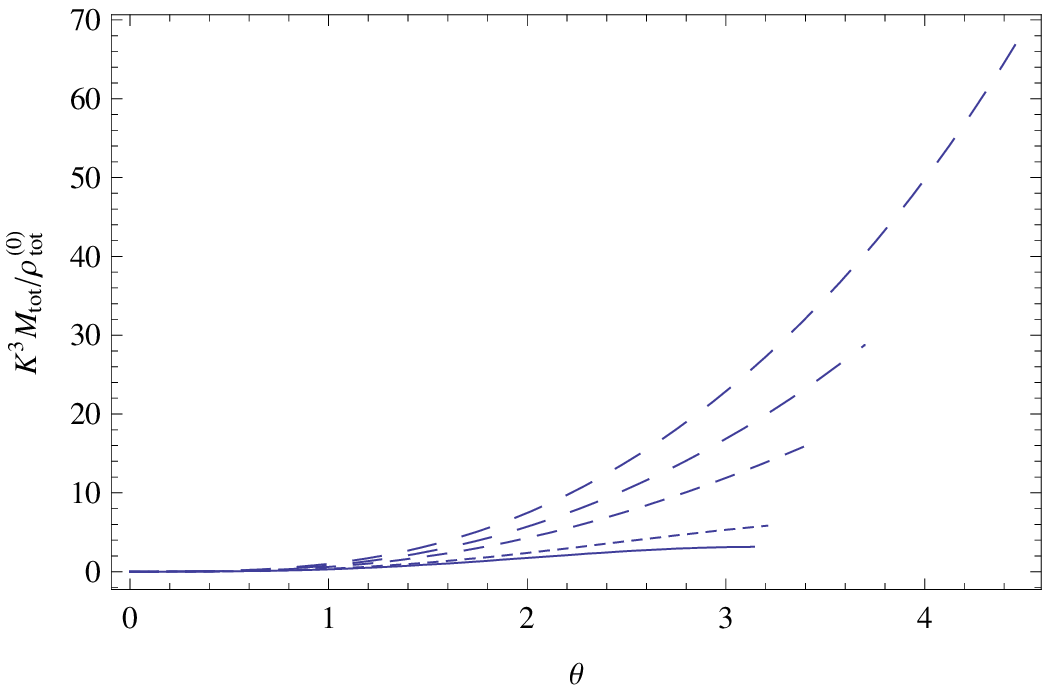}
\caption{Radial coordinate dependence of the total mass profile of the finite temperature dark matter halo, for different values of the temperature: $T=0$ (solid curve), $T/T_{BEC}=0.2$ (dotted curve), $T/T_{BEC}=0.4$ (dashed curve), $T/T_{BEC}=0.5$ (long dashed curve), and $T/T_{BEC}=0.6$ (ultra-long dashed curve). For all the curves $\alpha \left(a/m^{1/3}\right)=-5000\;{\rm cm/g^{1/3}}$. The Bose-Einstein transition density is taken as $\rho _{BEC}=10^{-21}\;{\rm g/cm^3}$, while the central density of the dark matter halo is $\rho _{tot}^{(0)}=10^{-24}\;{\rm g/cm^3}$.  The dimensionless radial coordinate $\theta $ is defined as $\theta =K(T)r$.  }
\label{FIG2}
\end{figure}

\begin{figure}[!ht]
\includegraphics[width=0.98\linewidth]{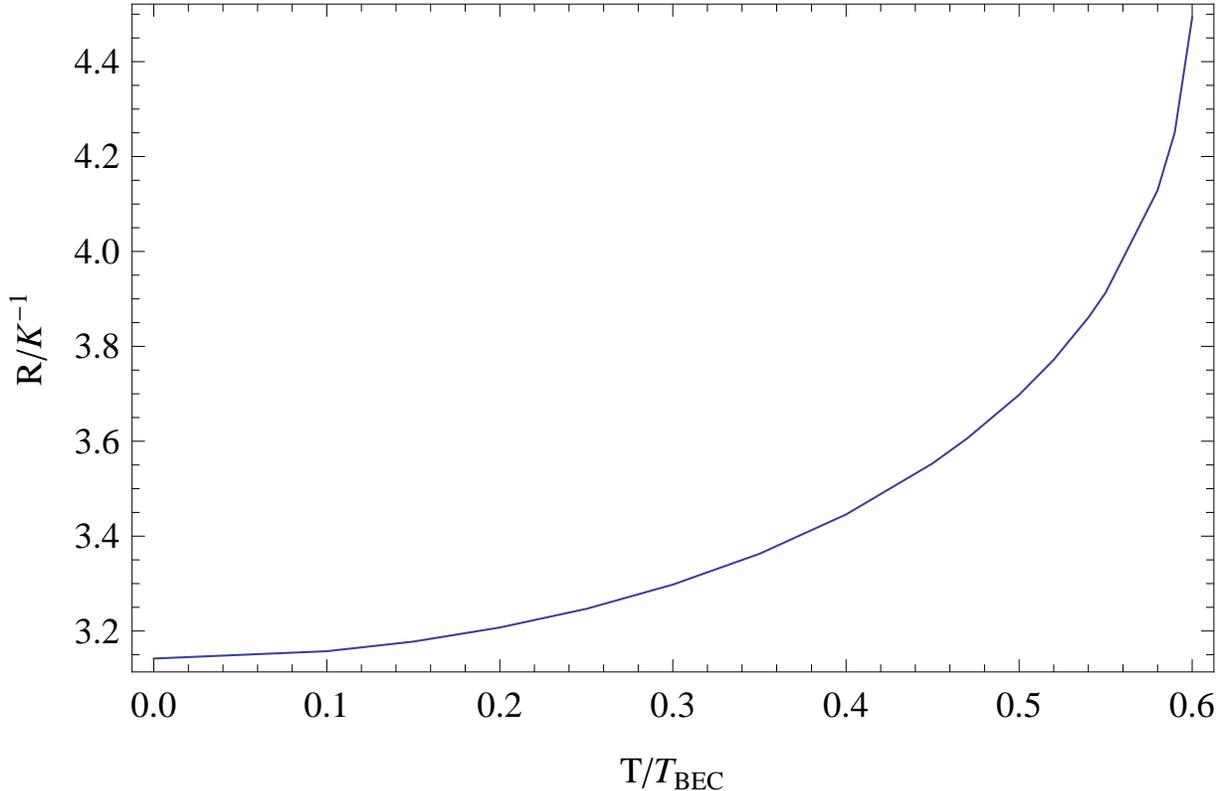}
\caption{Radius-temperature relation for the finite temperature dark matter halos. The values of the physical parameters are $\alpha \left(a/m^{1/3}\right)=-5000\;{\rm cm/g^{1/3}}$, $\rho _{BEC}=10^{-21}\;{\rm g/cm^3}$, while the central density of the dark matter halo is $\rho _{tot}^{(0)}=10^{-24}\;{\rm g/cm^3}$. }
\label{FIG3}
\end{figure}

The tangential velocity of test particles in stable circular orbits in the  finite temperature dark matter halos can be obtained as $v_{tg}^2=GM_{tot}/r$. The variations with respect to the dimensionless radial coordinate $\theta =K(T)r$ of the dark matter halo density and mass are presented, for different values of the temperature, in Figs.~\ref{FIG1} and \ref{FIG2}, respectively, while the temperature variation of the radius of the dark halo is shown in Fig.~\ref{FIG3}.

\section{Discussions and final remarks}\label{4}

In the present paper we have analyzed the effects of the finite temperature on the Bose-Einstein dark
matter halos. We have obtained the equations describing the condensate and the non-condensate component of the dark matter at finite temperatures $T>0$,  and we have derived the density profile,   the
mass distribution and the radius of the static finite temperature configuration in thermodynamic equilibrium. To obtain the description of the finite temperature dark matter halos we have considered an analytical approach, in which we have ignored the anomalous density term, and used a phenomenological description for the thermal excitations and non-condensed particles. This approach, despite its simplicity, gives a very clear physical description of the finite temperature properties of the dark matter halos.

We have explicitly determined the temperature dependence of the density profile of the dark matter halo, given by Eq.~(\ref{prof}).  As one can immediately see, if $T<<T_{BEC}$, the zero temperature density profile can give an excellent description of the condensate in thermodynamic equilibrium with a thermal cloud. The effect of the presence of the thermal cloud results in the modification of the central density of the system \cite{Ben}, which consists now from the sum of the central density of the condensate and of the thermal cloud. However, in order to see if the approximation of the zero temperature indeed works, one have to estimate the temperature at which the Bose-Einstein condensation took place, and the corresponding density of the Universe.

We assume that in the early stages of the evolution of the Universe dark matter consisted of bosonic particles of mass $m$ and temperature $T$, originating in equilibrium, and decoupling at a temperature $T_D$ or chemical potential $\mu _D >>m$. By assuming that the dark matter forms an isotropic gas of particles in kinetic equilibrium, the pressure of the non-condensed gas can be obtained as \cite{Har}
 \begin{equation}\label{pres1}
P=\rho _{DM }c^2\sigma _V^2,
\end{equation}
where $\rho _{DM }$ is the density of the dark matter particles before condensation, $\sigma _V^2=\langle \vec{v}^2 \rangle /3c^2$, and $\langle \vec{v}^2 \rangle$ is the average squared velocity of the particles. $\sigma _V$ is the one-dimensional velocity dispersion. By taking into account the continuity of the pressure during the condensation phase one can determine the transition density $\rho _{BEC}$. The numerical values of the physical parameters at the transition point depends on three unknown parameters, the dark matter particle mass $m$, the scattering length $a$, and the dark matter particles velocity dispersion $\sigma _V$, respectively. By assuming a typical mass of the dark matter particle of the order of 1 eV (1 eV = $1.78\times 10^{-33}$ g), a typical scattering length of the order of $10^{-10}$ cm, and a mean velocity square of the order of $\langle\vec{v}^2\rangle=81\times 10^{14}\;{\rm cm^2/s^2}$, the critical transition density can be written as \cite{Har}
\be
\rho _{BEC}=3.868\times 10^{-21}\left(\frac{\sigma _V^2}{3\times 10^{-6}}\right)\times
\left(\frac{m}{10^{-33}\;{\rm g}}\right)^3\left(\frac{a}{10^{-10}\;{\rm cm}}\right)^{-1}\;{\rm g/cm^3}.
\ee

The critical temperature at the moment of Bose-Einstein condensate transition is given by \cite{Har}
\be
T_{BEC}\approx 6.57089\times10^3\times\left(\frac{m}{10^{-33}\;{\rm g}}\right)^{1/3}\times
\left(\frac{\sigma _V^2}{3\times 10^{-6}}\right)^{2/3}\left(\frac{a}{10^{-10}\;{\rm cm}}\right)^{-2/3}\;K.
\ee

For the critical redshift we obtain the value \cite{Har}
 \begin{eqnarray}
1+z_{cr}&= &1.224\times10^{3}\times\left(\frac{m}{10^{-33}\;{\rm g}}\right)^{\left(1+\sigma _V^2\right)}\times \nonumber\\
&&\left(\frac{\sigma _V^2}{3\times 10^{-6}}\right)^{1/3\left(1+\sigma _V^2\right)}\left(\frac{a}{10^{-10}\;{\rm cm}}\right)^{-1/3\left(1+\sigma _V^2\right)}.
\end{eqnarray}

In obtaining these values for the critical for $\rho _{BEC}$, $T_{BEC}$ and $z_{cr}$ we have implicitly assumed that the Bose-Einstein condensation process took place relatively late in the history of the Universe, at a redshift $z_{cr}\approx 1000$. For a discussion of the possibilities of a Bose-Einstein condensation in the very early Universe see \cite{Bo}. By assuming that dark matter is cold, and in thermal equilibrium with the cosmic microwave background, it follows that the thermal contribution to the central density of the condensate is at least four orders of magnitude smaller than the central density of the dark matter, which may be of the order of $10^{-24}$ g/cm$^3$. Hence these simple qualitative estimation also show that for low redshift galaxies the general form of the density profile of the Bose-Einstein condensate dark matter can be described within an excellent approximation by the zero-temperature profile, $\rho =\rho _c\sin(Kr)/Kr$, even if we assume that the Bose-Einstein condensation of dark matter took place relatively late during the cosmological evolution of the Universe.

The zero-temperature approximation also allows us to make an estimate of the dark matter particle mass.   At the boundary of the dark matter
distribution $\rho (R)=0$, giving the condition $KR=\pi $,
which fixes the radius of the condensate dark matter halo as $R=\pi
\sqrt{\hbar ^{2}a/Gm^{3}}$.
The total mass of the condensate dark matter halo $M$ can be obtained as
$M=4\pi ^2\left(\hbar ^2a/Gm^3\right)^{3/2}\rho _c=4R^3\rho _c/\pi $,
giving for the mean value $<\rho >$ of the condensate density the expression $<\rho >=3\rho _c/\pi ^2$. The mass of the particle in the condensate is given by \cite{BoHa07}
\be
 m =\left( \frac{\pi ^{2}\hbar ^{2}a}{GR^{2}}\right) ^{1/3}\approx
6.73\times 10^{-2}\times
\left[ a\left( {\rm fm}\right) \right] ^{1/3}%
\left[ R\;{\rm (kpc)}\right] ^{-2/3}\;{\rm eV}.
\ee
For $a\approx
1 $ fm and $R\approx 10$ kpc, the typical mass of the condensate particle is of the order of $m\approx 14$
meV. For $a\approx 10^{6}$ fm, corresponding
to the values of $a$ observed in terrestrial laboratory experiments, $%
m\approx 1.44$ eV. These values are consistent with the limit $%
m<1.87$ eV obtained for the mass of the condensate particle from
cosmological considerations \cite{Bo}.

The properties of dark matter can be obtained observationally from the study of  the collisions between clusters
of galaxies, like the bullet cluster (1E 0657-56) and the baby bullet (MACSJ0025-12).
From these studies one can obtain constraints on the
properties of dark matter, such as its interaction cross-section
with baryonic matter and the dark matter-dark matter self-interaction cross section. If the ratio $\sigma _m=\sigma /m$ of the self-interaction cross section $\sigma =4\pi a^2$ and of the dark matter particle mass $m$ is known from observations, then the mass of the dark matter particle in the Bose-Einstein condensate can be obtained as
\begin{equation}
m=\left(\frac{\pi ^{3/2}\hbar ^2}{2G}\frac{\sqrt{\sigma _m}}{R^2}\right)^{2/5}.
\end{equation}

By comparing results from X-ray, strong lensing, weak lensing, and optical observations with numerical simulations
of the merging galaxy cluster 1E 0657-56 (the Bullet cluster), an upper limit (68 \% confidence) for $\sigma _m$ of the order of $\sigma _m<1.25\;{\rm cm^2/g}$  was obtained in \cite{Bul}. By adopting for $\sigma _m$ a value of $\sigma _m=1.25\;{\rm cm^2/g}$, we obtain for the mass of the dark matter particle an upper limit of the order
\begin{eqnarray}
m&<&3.1933\times10^{-37}\left(\frac{R}{10\;{\rm kpc}}\right)^{-4/5}\left(\frac{\sigma _m}{1.25\;{\rm cm^2/g}}\right)^{1/5}\;{\rm g}=\nonumber\\
&&0.1791\times\left(\frac{R}{10\;{\rm kpc}}\right)^{-4/5}\left(\frac{\sigma _m}{1.25\;{\rm cm^2/g}}\right)^{1/5}\;{\rm meV}.
\end{eqnarray}
By using this value of the particle mass we can estimate the scattering length $a$ as
\be
a<\sqrt{\frac{\sigma _m\times m}{4\pi }}=1.7827\times 10^{-19}\;{\rm cm}=1.7827\times 10^{-6}\;{\rm fm}.
\ee

This value of the scattering length $a$, obtained from the observations of the Bullet cluster 1E 0657-56 is much smaller than the value of $a=10^4-10^6$ fm corresponding to the BEC's obtained in laboratory terrestrial experiments \cite{exp}-\cite{Sr}.

If Bose-Einstein condensate dark matter is formed of Cold Dark Matter elementary particles, their superfluidity properties may distinguish them from other dark matter candidates due to the presence of quantum vortices. The presence of vortices in the central part of the BEC dark matter halos with their empty core change the gravitational coupling with baryons and contribute to a nearly smooth dark matter distribution in the halo.
The vortex stability condition is satisfied when the rotating halo has enough angular momentum to sustain more than one vortex like formation of vortex lattice due to rotation of spiral galaxies. Because of this rotation vortices manifesting as density holes appear in the condensate with quantized circulation,
$\Gamma = \oint_{C} \vec{v} \cdot d\vec{r} = l \left( h/m \right)$,
where $\Gamma$ is the circulation, $C$ is any contour around a vortex, and $l$ is the topological charge \cite{Y_M}.
$\Gamma$ is quantized in units of $h/m$. When $l=0$, there is no vortex. A rotating environment means that $l=1$ with one stable vortex or for the case of $l>1$ unstable vortices, which decay to the case having $l=1$.
Thus, rotation of a self-gravitating dark matter condensate is a source for quantized vortices, which evolve towards a vortex lattice. Its number density varies as $n_{v} \propto r^{-1}$, and suggests a flat velocity profile.

In a recent paper \cite{Slep}, the authors claim to have ruled out the Bose-Einstein condensate dark matter model. In order to support this conclusion an equation of state for the finite temperature condensate is derived. The isothermal equation of state obtained in \cite{Slep}  indicate a Bose-Einstein condensed core surrounded by a non-degenerate envelope, with an abrupt density drop marking the boundary between the two phases. By analyzing the behavior of the galactic rotation curves the authors conclude that since such a behavior is not confirmed observationally, the Bose-Einstein dark matter condensate model can be ruled out. The detailed analysis of the finite temperature effects performed in the present paper has shown that once the thermal effects are correctly taken into account, such a conclusion is incorrect. The main physical process missing in the analysis in \cite{Slep} is the requirement of a
static diffusive equilibrium between the cloud and the condensate, which forces the thermal cloud to follow the density distribution of the condensate. Moreover, since the temperature of the dark matter in low redshift galaxies is much smaller than the transition temperature, thermal effects do not play any significant role in the description of the dark matter halo density profiles, a result well-established in the condensed matter literature \cite{Za99,Prouk, Gr}. This situation is very similar to the case of other astrophysical systems having a pure quantum nature, like, for example, white dwarfs and neutron stars. Quantum statistical systems in which the thermal energy is much lower that the transition energy can be described with a high level of precision by the zero temperature limit. For example, the maximum mass and the physical properties of the white dwarfs and neutron stars (including the Chandrasekhar mass limit) are obtained for the $T=0$ case, even that the temperature of the stars may be of the order of $T=10^6-10^8$ K. But the $T=0$ approximation gives a very good description of the star because the thermal energy of the star is much lower than the Fermi energy of the electrons or neutrons. The situation for the dark matter halos is similar-if their temperature is much lower than the transition temperature, the $T=0$ approximation gives an excellent description of the system.

In conclusion, in the present paper we have shown that for temperatures much lower than the Bose-Einstein transition temperature, $0<T<0.5T_{BEC}$ the finite temperature effects on the density profiles of the condensed dark matter halos can be neglected. The zero-temperature density profiles gives an excellent description of the observed properties of the rotation curves \cite{BoHa07}. However, for $T\geq 0.4T_{BEC}$, finite temperature effects can play an important role in determining both the static and dynamic properties of the condensate. Hence finite temperature effects can play an important role in a cosmological context and for the analysis of the cosmological dynamics after the condensation.

\acknowledgments

We would like to thank to the anonymous referee for comments and suggestions that helped us to significantly improve our manuscript. TH is supported by an RGC grant of the government of the Hong Kong SAR.

\end{document}